\documentclass[twocolumn]{aastex62}

\def\mosfit{\texttt{mosfit}}

\def\hst{\textit{HST}}

\newcommand{\kms}{km\,s$^{-1}$}
\def\cmg{cm$^2$\,g$^{-1}$}
\def\ergs{erg\,s$^{-1}$}
\def\magd{\,mag\,(100\,d)$^{-1}$}

\def\M{M$_{\odot}$}

\def\Ha{H{$\alpha$}}

\def\Ni{$^{56}$Ni}
\def\Co{$^{56}$Co}

\def\Mej{$M_\mathrm{ej}$}

\def\vej{$v_\mathrm{ej}$}

\def\ergs{erg\,s$^{-1}$}

\def\ca{[\ion{Ca}{2}]\,$\lambda$7300}
\def\o{[\ion{O}{1}]\,$\lambda$6300}

\def\oi{\ion{O}{1}\,$\lambda$7774}

\shorttitle{1000 days of SN\,2015bn}
\shortauthors{Nicholl et al.}

\begin{document}

\title{One thousand days of SN\,2015bn: \textit{HST} imaging shows a light curve flattening consistent with magnetar predictions}

\correspondingauthor{Matt Nicholl}
\email{mrn@roe.ac.uk}

\author[0000-0002-2555-3192]{Matt Nicholl}
\affil{Harvard-Smithsonian Center for Astrophysics, 60 Garden Street, Cambridge, MA, 02138, USA}
\affil{Institute for Astronomy, University of Edinburgh, Royal Observatory, Blackford Hill, Edinburgh EH9 3HJ, UK}

\author{Peter K.~Blanchard}
\affil{Harvard-Smithsonian Center for Astrophysics, 60 Garden Street, Cambridge, MA, 02138, USA}

\author{Edo Berger}
\affil{Harvard-Smithsonian Center for Astrophysics, 60 Garden Street, Cambridge, MA, 02138, USA}

\author{Kate D.~Alexander}
\affil{Harvard-Smithsonian Center for Astrophysics, 60 Garden Street, Cambridge, MA, 02138, USA}

\author{Brian D.~Metzger}
\affil{Department of Physics and Columbia Astrophysics Laboratory, Columbia University, New York, NY 10027, USA}

\author{Kornpob Bhirombhakdi}
\affil{Astrophysical Institute, Department of Physics and Astronomy, 251B Clippinger Lab, Ohio University, Athens, OH 45701, USA}

\author{Ryan Chornock}
\affil{Astrophysical Institute, Department of Physics and Astronomy, 251B Clippinger Lab, Ohio University, Athens, OH 45701, USA}

\author{Deanne Coppejans}
\affil{Center for Interdisciplinary Exploration and Research in Astrophysics (CIERA) and Department of Physics and Astronomy, Northwestern University, Evanston, IL 60208, USA}

\author{Sebastian Gomez}
\affil{Harvard-Smithsonian Center for Astrophysics, 60 Garden Street, Cambridge, MA, 02138, USA}

\author{Ben Margalit}
\affil{Department of Physics and Columbia Astrophysics Laboratory, Columbia University, New York, NY 10027, USA}

\author{Raffaella Margutti}
\affil{Center for Interdisciplinary Exploration and Research in Astrophysics (CIERA) and Department of Physics and Astronomy, Northwestern University, Evanston, IL 60208, USA}

\author{Giacomo Terreran}
\affil{Center for Interdisciplinary Exploration and Research in Astrophysics (CIERA) and Department of Physics and Astronomy, Northwestern University, Evanston, IL 60208, USA}

\begin{abstract}

We present the first observations of a Type I superluminous supernova (SLSN) at $\gtrsim1000$ days after maximum light. We observed SN\,2015bn using the \textit{Hubble Space Telescope} Advanced Camera for Surveys in the F475W, F625W and F775W filters at 721 days and 1068 days. SN\,2015bn is clearly detected and resolved from its compact host, allowing reliable photometry. A galaxy template constructed from these data further enables us to isolate the SLSN flux in deep ground-based imaging. We measure a light curve decline rate at $>700$ days of $0.19\pm0.03$\,\magd, much shallower than the earlier evolution, and slower than previous SLSNe (at any phase) or the decay rate of \Co. Neither additional radioactive isotopes nor a light echo can consistently account for the slow decline. A spectrum at 1083 days shows the same \o\ and \ca\ lines as seen at $\sim300-400$  days, with no new features to indicate strong circumstellar interaction. Radio limits with the Very Large Array rule out an extended wind for mass-loss rates $10^{-2.7}\lesssim\dot{M}/v_{10}\lesssim10^{-1.1}$\,\M\,yr$^{-1}$ (where $v_{10}$ is the wind velocity in units of 10\,\kms). The optical light curve is consistent with $L\propto t^{-4}$, which we show is expected for magnetar spin-down with inefficient trapping; furthermore, the evolution matches predictions from earlier magnetar model fits. The opacity to magnetar radiation is constrained at $\sim0.01$\,\cmg, consistent with photon-matter pair-production over a broad $\sim$GeV--TeV range. This suggests the magnetar spectral energy distribution, and hence the `missing energy' leaking from the ejecta, may peak in this range.

\end{abstract}


\section{Introduction} \label{sec:intro}

Hydrogen-poor superluminous supernovae (Type I SLSNe; here simply SLSNe) are a rare class of massive star explosions with typical peak absolute magnitudes $M\sim-21$\,mag \citep{qui2011,chom2011}. Despite intense observational and theoretical study, the energy source underlying their light curves has remained uncertain \citep[e.g.][]{moriya2018b}.

\begin{figure*}
\includegraphics[width=17cm]{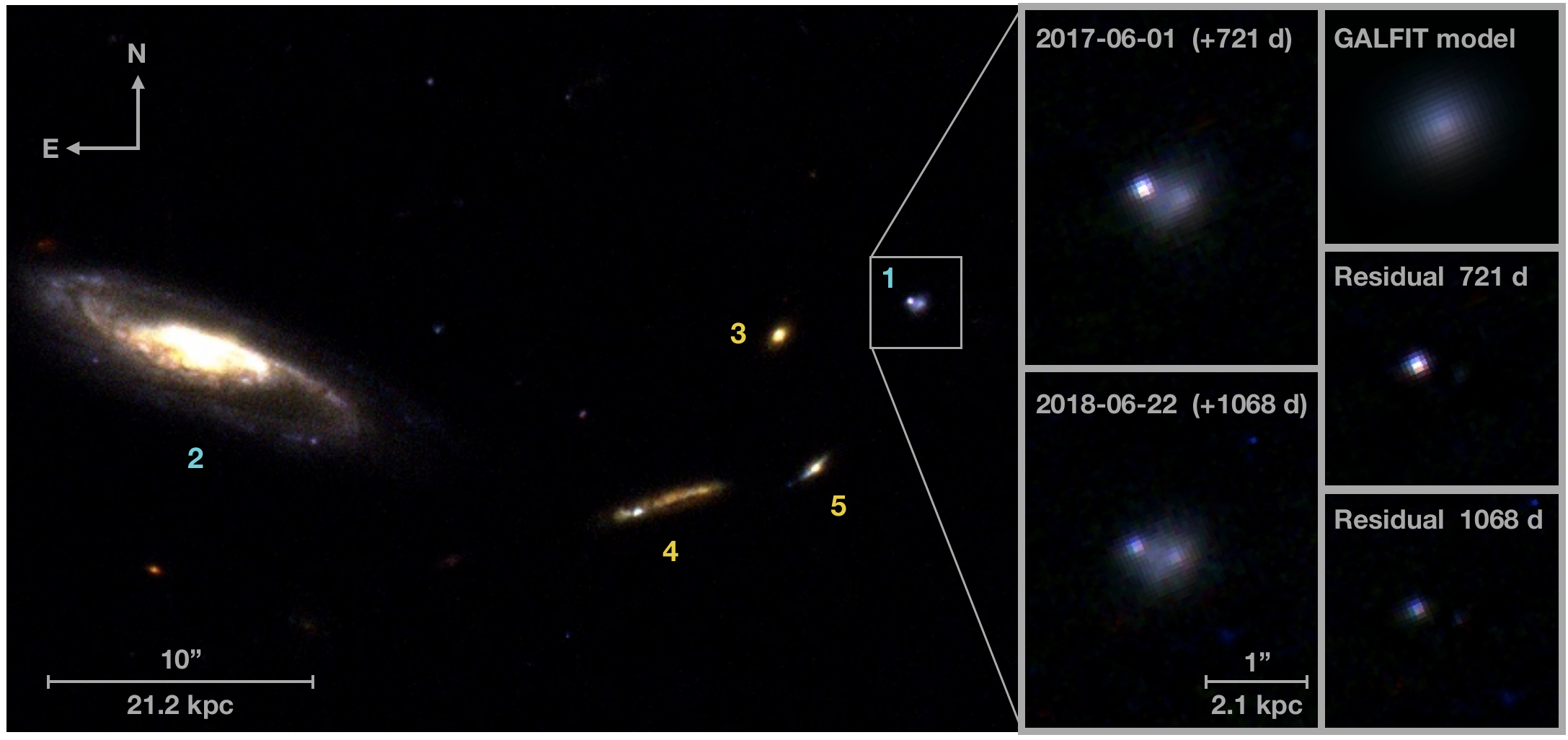}
\centering
\caption{\hst\ imaging of SN\,2015bn at 721--1068 rest-frame days after maximum. The image on the left is a $gri$ three-colour composite of the earlier epoch. The host galaxy (labelled 1) and the large spiral (2) are at consistent redshifts, $z\approx0.11$, with separation and relative magnitudes comparable to the SMC and Milky Way. The other three sources (3--5) are background galaxies at $z\approx0.35$. Panels on the right show a zoom-in around SN\,2015bn and the subtraction of a galaxy model with \texttt{galfit}. The SN is clearly detected, fading by a factor $\sim2$ between observations.} 
\label{fig:hst}
\end{figure*}

Normal stripped envelope SNe are powered by $\sim\!\mathrm{few}\times0.1$\,\M\ of synthesized \Ni\ \citep[e.g.][]{dro2011}, whereas SLSNe would require several solar masses if that was the primary energy source. Such a large \Ni\ mass conflicts with the early light curves \citep{nic2013}, late-time limits \citep{bla2018}, and with spectra \citep{des2012,jer2017a,nic2018}, but a number of SLSNe do fade at a rate that resembles the decay of \Co, the daughter nucleus of \Ni\ \citep{gal2009,dec2018}.

The most popular model for SLSNe is the spin-down of a millisecond magnetar with magnetic field $B\gtrsim10^{13}$\,G \citep{kas2010}. While this reproduces most SLSN observables \citep{ins2013,nic2017c}, a `smoking gun' has proven elusive; thus competing models, such as ejecta interacting with a circumstellar medium (CSM), remain competitive. It was hoped that a magnetar engine could drive an X-ray breakout months after the explosion \citep{met2014}, but this has not been detected \citep{margutti2017,ins2017}, and more recent (and realistic) models predict that breakouts should be rare \citep{margalit2018}.

A more robust test for the magnetar engine comes from the late-time light curve. The spin-down luminosity ultimately follows a power-law, $L\propto t^{-\alpha}$, so eventually the decline should become shallower than \Co\ decay, which follows an exponential (half-life\,$\approx77$ days). While many SLSN light curves have been observed to flatten at late times, the spin-down rate can remain within a factor of a few of \Co\ decay for hundreds of days \citep{ins2013,moriya2017}, and most SLSNe are too distant to follow to such late phases.

In this Letter, we report the first detections of a Type I SLSN at $\gtrsim\!1000$ days after maximum light. SN\,2015bn is a slowly-evolving SLSN at $z=0.1136$, and has been extensively studied at earlier times \citep{nic2016b,nic2016c,ins2016,jer2017a,leloudas2017}. New imaging with the \textit{Hubble Space Telescope} (\textit{HST}) and Magellan reveals a marked flattening in the light curve after $\sim500$ days, consistent with a power law, and a decline rate that is now significantly slower than \Co\ decay. Spectroscopy and radio follow-up show no signs of circumstellar interaction. After eliminating several other possibilities, we argue that this is best interpreted as the signature of a magnetar engine.

\begin{figure*}
\includegraphics[width=12cm]{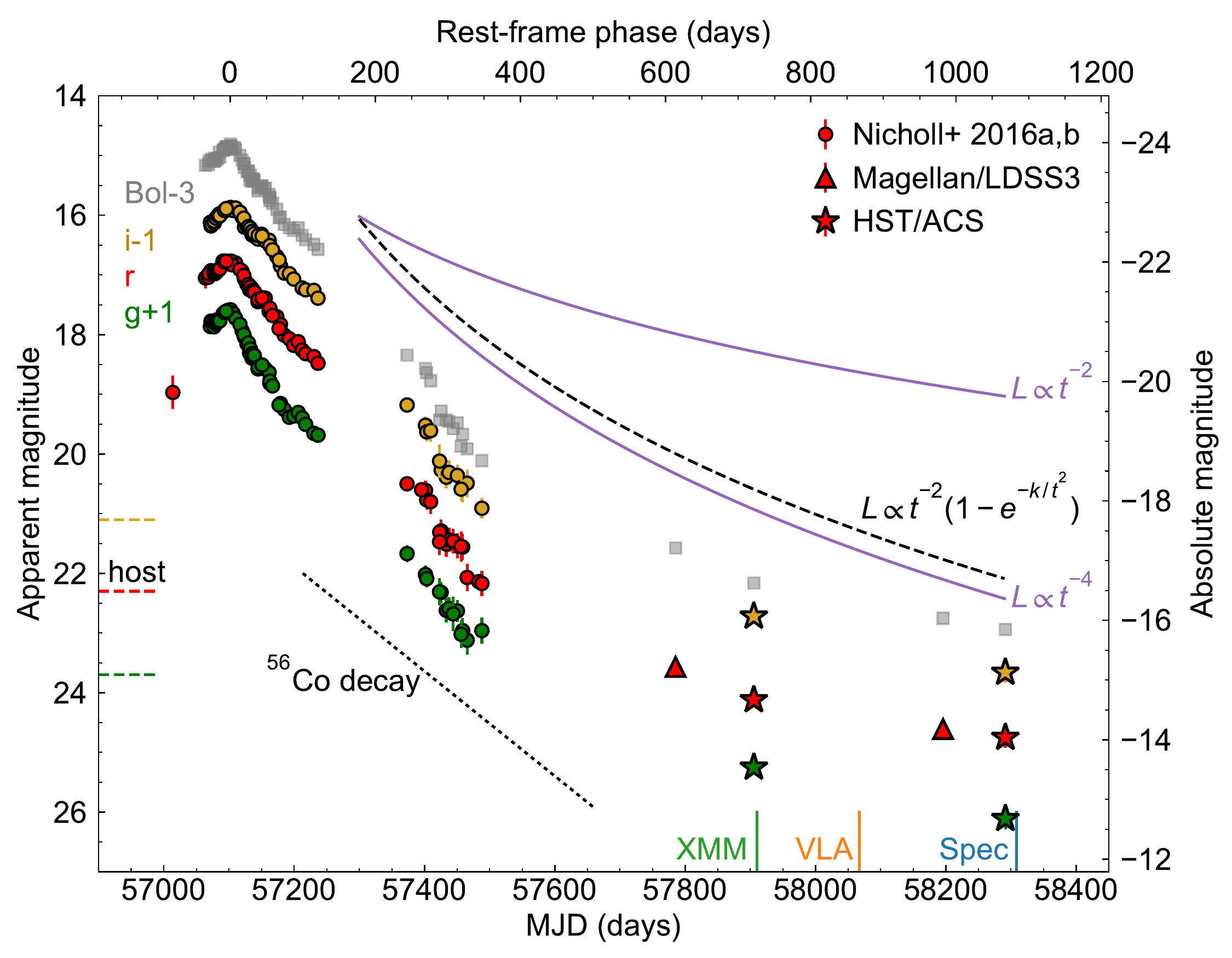}
\centering
\caption{Light curve of SN\,2015bn. The points labelled `Bol' are the pseudobolometric magnitudes obtained by integrating the $gri$ flux. The absolute scale on the right axis assumes a constant $K$-correction of $-2.5 \log(1+z)$. Host magnitudes are from SDSS. The decline rate at $\gtrsim500$ days is much slower than before, and clearly shallower than \Co\ decay (dotted line). The post-maximum light curves are broadly consistent with a power-law, $L \propto t^{-4}$, equivalent to magnetar spin-down with incomplete thermalisation (dashed line). Dates of spectroscopic and radio data (this work), and X-ray data \citep{bhirombhakdi2018}, are marked, as are host galaxy magnitudes (offset to match light curves).} 
\label{fig:lc}
\end{figure*}

\section{Observations}
\label{sec:data}

\subsection{Optical imaging}

We imaged SN\,2015bn using the \hst\ Advanced Camera for Surveys Wide Field Channel\footnote{Program IDs: 14743,15252; PI: Nicholl} on 2017-06-01.4 and 2018-06-22.3 (all dates in UT), corresponding to 721 and 1068 days after maximum light in the rest-frame of SN\,2015bn.
Visits consisted of one orbit per filter in F475W, F625W and F775W, corresponding closely to $g$, $r$ and $i$ bands, where we expect most of the strong emission lines \citep{nic2016c,jer2017a,nic2018}. Each image contained four dithers in a standard box pattern.

We retrieved the fully processed and drizzled images from the Mikulski Archive for Space Telescopes. Figure \ref{fig:hst} shows the combined three-colour images. SN\,2015bn is clearly visible as a point source superimposed on its host galaxy. We removed the host using a galaxy model constructed with \texttt{galfit} \citep{peng2002}, fitting a S\'ersic profile while masking the pixels that were clearly dominated by SN\,2015bn. There were no significant differences between the fits obtained in the individual epochs. Subtracting the model from the \hst\ images resulted in a clean SN detection with minimal galaxy residuals, as shown in Figure \ref{fig:hst}. 
We then performed point-spread function (PSF) photometry with \texttt{daophot}, and applied standard zeropoints\footnote{http://www.stsci.edu/hst/acs/analysis/zeropoints}. We verified that the zeropoints were consistent between the two epochs (to within $<0.02$\,mag) using 16 stars from the Pan-STARRS Data Release 1 catalog \citep{flewelling2016}.

We obtained ground-based imaging on 2017-02-01.3 and 2018-03-18.7 using the Low Dispersion Survey Spectrograph 3 (LDSS3) on the 6.5-m Magellan Clay telescope. Each observation consisted of 10$\times$300\,s dithered $r$-band exposures, which we reduced in \texttt{pyraf}. From the ground, SN\,2015bn appears entirely blended with its (much brighter) host. Subtracting the \texttt{galfit} model derived from the \hst\ data, after convolving to the ground-based resolution using \texttt{hotpants}\footnote{https://github.com/acbecker/hotpants}, we isolated the SN light and performed PSF photometry, determining the zeropoints using the Pan-STARRS catalog.

Our photometry is plotted in Figure \ref{fig:lc}, along with earlier $g,r,i$ data from \citet{nic2016b,nic2016c}. The latest points are fainter than the peak by a factor $\approx\!1500$, but a flattening in the light curve beyond $\sim500$ days is immediately apparent. The new data have been submitted to the \textit{Open Supernova Catalog} \citep{guillochon2017}.

\subsection{SN spectroscopy}

We observed SN\,2015bn spectroscopically on 2018-07-08.9 (1083 rest-frame days after maximum) using LDSS3. The data were reduced in \texttt{pyraf}, with flux calibration achieved using a standard star. The spectrum is shown in Figure \ref{fig:spec}. The mean airmass during the observation was 1.6, and the spectrum redwards of $\sim7500$\AA\ is contaminated by noise residuals from sky subtraction.

Although the spectrum is dominated by the host, the strongest emission lines from SN\,2015bn appear to be visible above the galaxy light. We subtract a model for the host continuum \citep{nic2016b}, and compare to the most recent prior spectrum \citep[at 392 days after maximum;][]{nic2016c}, scaled to match the latest \hst\ observations. We find that the broad feature at 6300\,\AA\ is consistent with predictions for \o, while a tentative feature at 7300\,\AA\ matches \ca. This indicates that the lines have changed little, despite a gap of 691 days. Our new spectrum likely represents the oldest spectroscopic detection with respect to explosion for any SLSN: the normalized phase is $t/t_d=13.5$ in the terminology of \citet{nic2018}, where $t_d=80$ days is the decline timescale of the light curve.

\subsection{Galaxy spectroscopy}

We also obtained spectra of three galaxies that apparently neighbour SN\,2015bn (labelled 3--5 in Figure \ref{fig:hst}). 
We find they are a background group at $z=0.353$ unrelated to SN\,2015bn. The bright ($M_r\approx-21$) spiral galaxy (2) has a spectrum from the Sloan Digital Sky Survey Data Release 7 \citep{abazajian2009} that indicates $z=0.1118$, similar to SN\,2015bn.

The relative line-of-sight velocity between this galaxy and the SN host is $c\Delta z=540$\,\kms, while their projected separation is $\approx56$\,kpc. 
These values are similar to the Magellanic Clouds relative to the Milky Way, and the absolute magnitude ($M_r=-16.4$), physical size, star-formation rate and metallicity of the host \citep{nic2016b} are all similar to the SMC. Thus the host and its bright neighbour appear to be a close analogue of the MW-SMC system.

\citet{chen2017} found that the host of one SLSN, LSQ14mo, was in a likely interacting system with a projected separation of 15\,kpc, and proposed that this could increase the likelihood of SLSNe by triggering vigorous star formation.
The brightest and bluest pixels in our \hst\ images, likely corresponding to the highest star-formation rate, actually appear to be on the other side of the galaxy, though we cannot exclude comparable star-formation at the position of SN\,2015bn until the SLSN has completely faded.

\subsection{Radio observations}

We observed SN\,2015bn using the Karl G.~Jansky Very Large Array (VLA) in B configuration, on 2017-11-10 (867 rest-frame days after maximum)\footnote{Program ID: 17B-164; PI: Nicholl}. SN\,2015bn was not detected to $3\sigma$ limiting flux densities of 48\,$\mu$Jy in K band (21.8\,GHz) and 63\,$\mu$Jy in Ka band (33.5\,GHz).

\begin{figure}
\includegraphics[width=8.2cm]{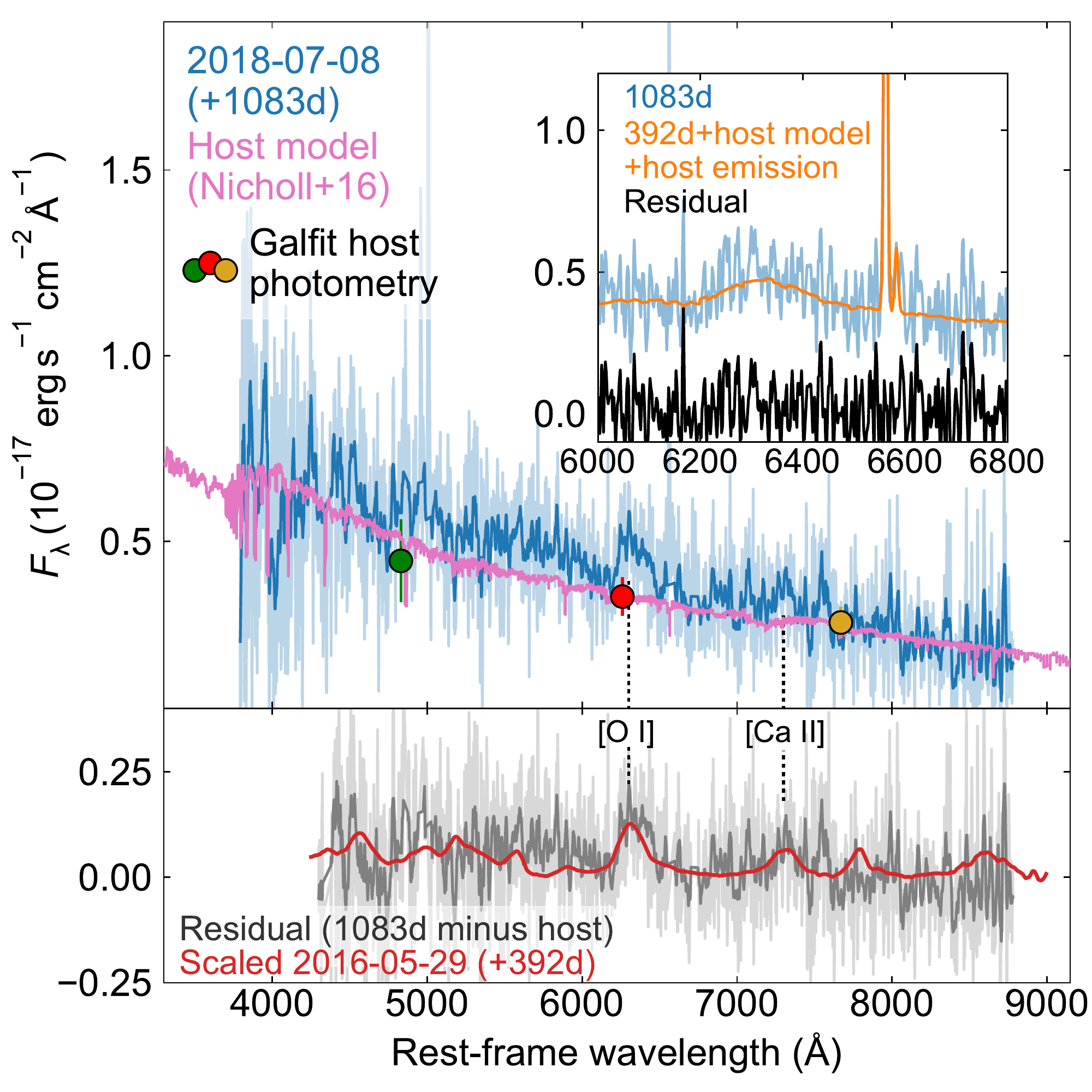}
\centering
\caption{Top: Spectrum of SN\,2015bn at 1083 days. The spectrum is dominated by galaxy light, and has been scaled to match the host photometry. We have smoothed the data with a Savitsky-Golay filter for clarity. We also plot the model host spectrum from \citet{nic2016b}.
Bottom: Host-subtracted spectrum. Broad features in the residuals match \o\ and possibly \ca. We compare to the spectrum of SN\,2015bn at 392 days \citep{nic2016c}, scaled to the latest magnitudes from \hst. The features in the new spectrum are consistent with the 392-day spectrum, supporting their identification as SN lines. Inset: Zoom-in around \o\ and \Ha. The 1083-day spectrum can be modelled as the sum of the host continuum model, the scaled 392-day spectrum, and Gaussian fits to the narrow \Ha\ and [\ion{N}{2}] data. Widths have been fixed at the instrumental resolution (8\,\AA). No broad component to \Ha, which would indicate circumstellar interaction \citep{yan2017}, is observed.} 
\label{fig:spec}
\end{figure}

\section{Analysis}
\label{sec:analysis}

The principal discovery from our observations is the shallow light curve beyond 500 days. The mean slope in $g,r,i$ between the \hst\ epochs is $0.19\pm0.03$\,\magd. Integrating the flux over these bands yields a similar pseudobolometric decline of $0.22\pm0.02$\,\magd. This is the slowest decline rate measured for any hydrogen-poor SLSN, and is significantly slower than the 1.43\,\magd\ during the first 400 days \citep{nic2016c}, and the \Co\ decay rate of 0.98\,\magd.

Since few SLSNe have deep observations at this phase, it is possible that others reach a similarly slow decline; however, the only other SLSN with photometry at a comparable phase, PTF10nmn, did not show a change in slope up to $\approx 700$ days \citep{dec2018}. The light curve of PS1-14bj appeared to reach to a slope flatter than \Co\ by around 400 days \citep{lun2016}, but further monitoring was not available to confirm this. We now examine possible causes of the flattening in SN\,2015bn.

\subsection{Light echo?}

Light echoes occur when light emitted earlier in the SN evolution is reflected into our line of sight by nearby dust sheets, giving an apparent luminosity boost after a light travel time.
For nearby SNe, this is readily identifiable through a change in the spatial emission profile, but at the distance of SN\,2015bn, 1 light year corresponds to only $\sim 10^{-4}$ arcseconds. An echo beginning $\sim 2$ years after explosion could roughly match the late-time brightness if it was $\approx 8$ magnitudes fainter than the light curve peak. 
\citet{lunnan2018b} recently detected the first light echo in a H-poor SLSN, iPTF16eh, via a \ion{Mg}{2} resonance line.

There are several issues with interpreting the behaviour of SN\,2015bn as an echo. First, the luminosity of an echo is expected to evolve as $t^{-1}$ \citep[e.g][]{graur2018a}, which is flatter than what we observe. 
Second, the spectrum is consistent with a typical SLSN nebular spectrum, whereas an echo should contain features from earlier phases, when the SN was brighter. However, we caution that the spectrum is noisy and dominated by host galaxy light.

Finally, dust is more efficient in reflecting blue light, which changes the observed colours. We measure $g-i=0.45\pm0.24$ at 1068 days, which is consistent with the colour at 300 days ($g-i=0.35\pm0.17$) but not with the peak ($g-i=-0.27 \pm 0.02$). A similar finding applies to $g-r$ and $r-i$. We therefore conclude that an echo cannot explain the slow evolution.

\subsection{Radioactive isotopes?}\label{sec:co}

Follow-up of nearby SNe at $\gtrsim900$ days has revealed evidence for the decay chain $^{57}$Ni$\rightarrow ^{57}$Co$\rightarrow ^{57}$Fe, in both core-collapse \citep{seitenzahl2014} and Type Ia SNe \citep{shappee2017,graur2018a}. While the relative abundance of $^{57}$Ni is typically low ($^{57}$Ni/\Ni\,$\lesssim 0.05$), the long lifetime of $^{57}$Co (half-life\,$\approx272$ days) means that it eventually comes to dominate over \Co.

The decay slope for $^{57}$Co is 0.28\magd, which is comparable to our light curve, but still somewhat faster. A small contribution from the slower reaction $^{55}$Fe$\rightarrow^{55}$Mn (half-life\,$\approx1000$ days) could help to mitigate this. \citet{seitenzahl2014} also looked for signatures of $^{60}$Co and  $^{44}$Ti in SN\,1987A, but the half-lives of these species are too long (5--60 years) to be relevant to SN\,2015bn yet.

The more significant problem for this scenario is that the pseudobolometric luminosity of SN\,2015bn at 900 days is $10^{40.8}$\,\ergs, i.e.~400--4000 times greater than SNe Ia at the same phase \citep{graur2018a}. The required $^{57}$Co mass is $\gtrsim 6$\,\M. We are not aware of any explosion model capable of producing this; even the most massive pair-instability models from \citet{heg2002} synthesize an order of magnitude less $^{57}$Co (while making 40\,\M\ of \Co). For a solar ratio of $^{57}$Co/\Co\,$=0.023$ \citep{lod2003}, the implied \Co\ mass would be $>260$\,\M.

\subsection{Circumstellar interaction?}\label{sec:csm}

Assuming a velocity of $\sim 7000$\,\kms\ \citep{nic2016b,nic2016c}, the ejecta expand to a radius $\approx6\times10^{16}$\,cm within 1000 days, and the fastest ejecta likely reach $\sim10^{17}$\,cm. \citet{yan2017} found that up to $\sim15$\% of SLSNe encounter hydrogen-rich CSM at $\sim10^{16}$\,cm, as indicated by the sudden appearance of broad hydrogen emission lines in their spectra, while \citet{lun2018} identified a circumstellar shell at $\gtrsim10^{17}$\,cm around iPTF16eh from its light echo. 
Thus interaction with a massive CSM at a similar radius could be a plausible luminosity source for SN\,2015bn. 

\begin{figure}
\includegraphics[width=8.2cm]{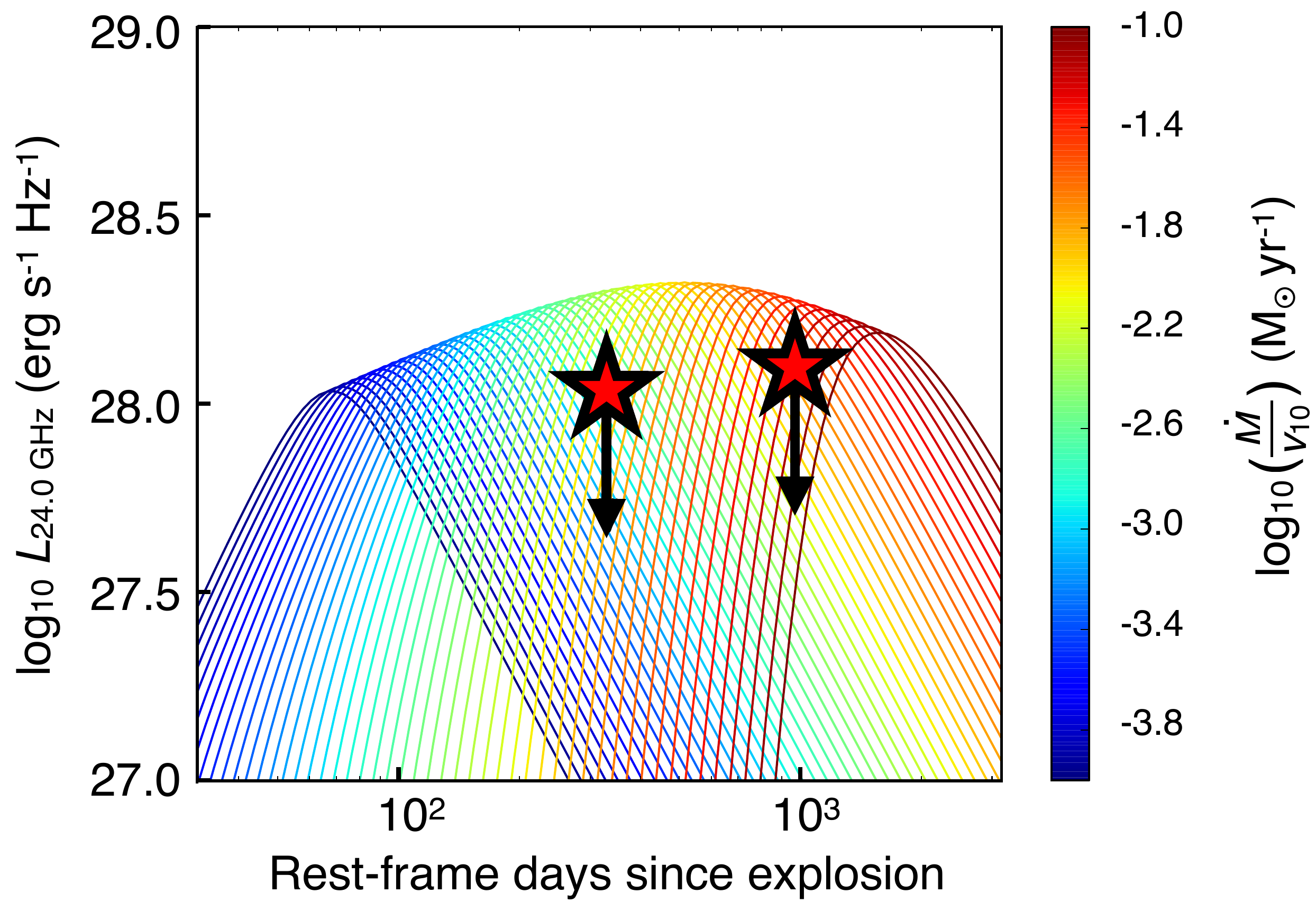}
\centering
\caption{VLA upper limits on the 24\,GHz (rest-frame) emission from SN\,2015bn. The earlier limit is from \citet{nic2016b}. Overplotted are models from \citet{kam2016} predicting the radio emission from a SN shock expanding into a circumstellar wind (density $\propto r^{-2}$). The VLA non-detections rule out extended winds corresponding to mass-loss rates of $10^{-2.7}\lesssim\dot{M}/v_{10}\lesssim10^{-1.1}$\,\M\,yr$^{-1}$.} 
\label{fig:vla}
\end{figure}

However, none of the interacting events in \citet{yan2017} showed a shallow light curve resembling SN\,2015bn, though the interaction in those events occurred much earlier (100--250 days) when the SLSNe were $\sim 1-2$ orders of magnitude brighter.
\citet{chen2018} recently studied SN\,2017ens, another SLSN that developed strong and broad \Ha\ emission at $\gtrsim100$ days, finding its light curve was essentially flat at this phase.

We examine the \Ha\ region of our spectrum in Figure \ref{fig:spec}. We subtract a model consisting of the scaled 392-day spectrum and a linear host continuum, and fit the \Ha\ and [\ion{N}{2}] lines with Gaussian profiles. A satisfactory fit is obtained with the width fixed at the instrumental resolution; i.e.~the lines are unresolved, and no broad component is present above the level of the noise. The flux in \Ha\ is $1.6\times10^{-16}$\,erg\,s$^{-1}$\,cm$^{-2}$, consistent with host emission \citep{nic2016b}. 

While \Ha\ is generally the strongest line in SNe interacting with hydrogen-rich material, interaction with hydrogen-free material is more difficult to exclude. \citet{benami2014} detected narrow [\ion{O}{1}]\,$\lambda5577$ emission from SN\,2010mb, and proposed it was a signature of interaction. We do not observe this line in SN\,2015bn to a limit of $\lesssim4\times10^{37}$\,\ergs, which is $\sim 10-100$ times fainter than the line in SN\,2010mb up to one year after explosion. A possible caveat is that this line is only predicted to be strong at densities $>10^7$\,g\,cm$^{-3}$.

Late-onset interaction in other events has been interpreted as a collision with a detached shell, but a slow decline could also result from an extended dense wind.
Figure \ref{fig:vla} shows predicted radio emission for SNe interacting with winds of different densities \citep{kam2016}, which we compare to our VLA limit and an earlier limit from \citet{nic2016b}.
Parameterising the wind mass-loss rate as $\dot{M}/v_{10}$, where $v_{10}$ is the wind velocity in units of 10\,\kms, the combined limits at 1--3 years rule out winds with $10^{-2.7}\lesssim\dot{M}/v_{10}\lesssim10^{-1.1}$\,\M\,yr$^{-1}$.
For a typical Wolf-Rayet wind velocity $\sim1000$\,\kms, this corresponds to $10^{-4.7}\lesssim\dot{M}\lesssim10^{-3.1}$\,\M\,yr$^{-1}$, excluding a wind significantly more dense than those from SN Ic progenitors \citep[e.g.][]{ber2002,soderberg2006,dro2015}.

Comparing to models for the optical luminosity from \citet{nic2016b}, this rules out most of the parameter space where the light curve peak can be powered by interaction with a dense wind, and also disfavours this as the primary late-time power source.

\subsection{Magnetar spin-down?}\label{sec:mag}

The most popular model for SLSNe is a magnetar engine. At late times the engine power decays as $L \propto t^{-\alpha}$; a standard magnetic dipole has $\alpha=2$. A long-standing prediction is that SLSNe should eventually track this power-law. While many SLSN light curves have been observed to flatten with time, late observations have generally been either similar to \Co\ decay \citep{ins2013} or of insufficient signal-to-noise ratio \citep{lun2016} to make strong statements.

In Figure \ref{fig:lc}, we plot representative curves for $\alpha=2$ and $\alpha=4$. The best-fitting power-law at 200--1100 days has $\alpha \approx 3.8$, steeper than a standard dipole. However, it is expected that the energy available from spin-down is not completely thermalized at late-times; assuming this energy is injected primarily as high-energy photons, the optical depth in the expanding ejecta decreases with time as $\tau\propto t^{-2}$ \citep{wang2015,chen2015}. Including this `leakage' term gives
\begin{equation}
L\propto t^{-2}(1-e^{-kt^{-2}})
\approx kt^{-4},
\end{equation}
where $k=3\kappa_\gamma M_\mathrm{ej}/(4\pi v_\mathrm{ej}^2)$ is the trapping coefficient, \Mej\ and \vej\ are the mass and velocity of the ejecta and $\kappa_\gamma$ is the opacity to high-energy photons. The second equality comes from a Taylor expansion applicable at late times. Thus a realistic power-law is not $\alpha=2$, but rather $\alpha\approx 4$, close to what we observe.

\begin{figure}[t]
\includegraphics[width=8.2cm]{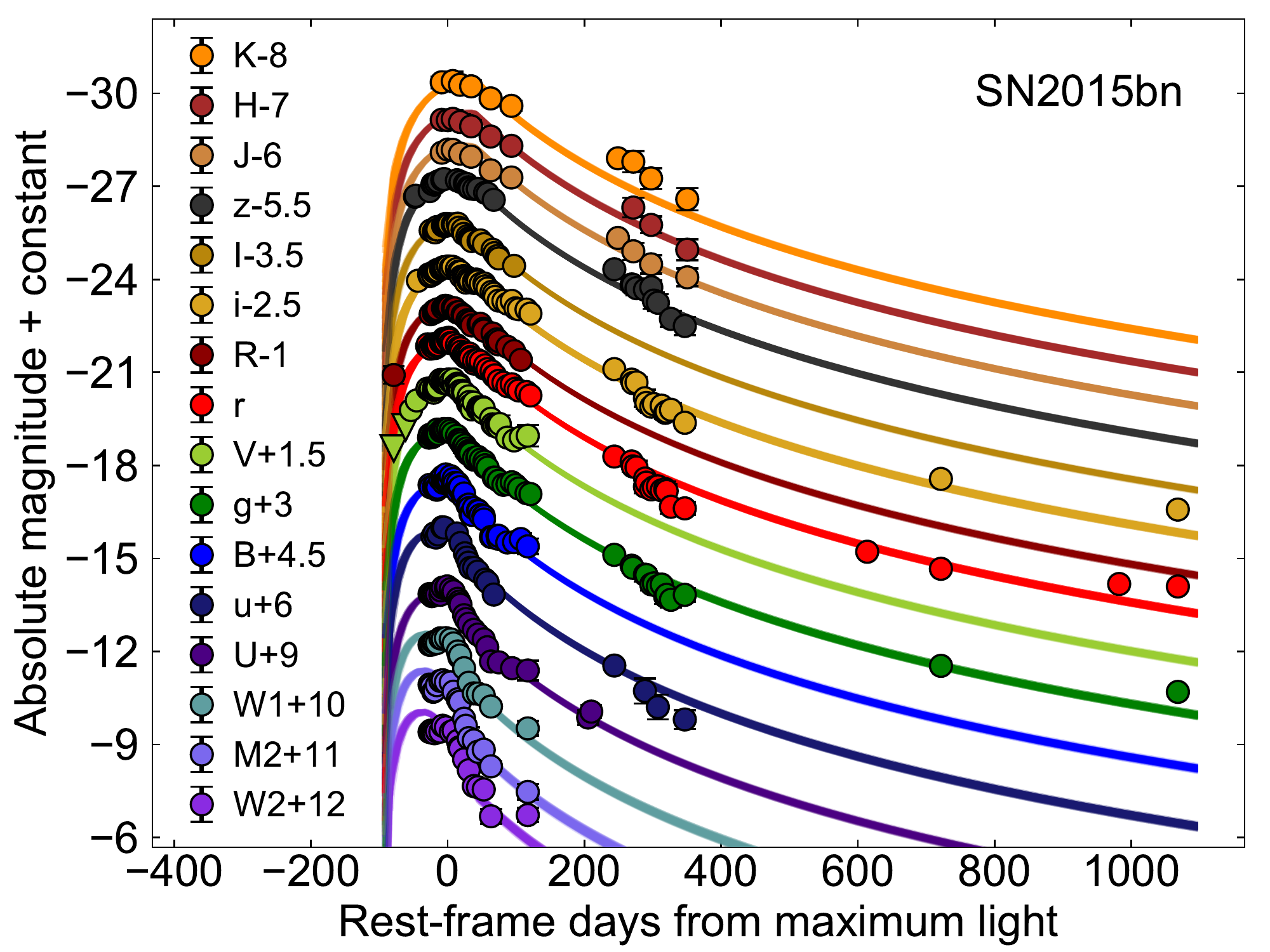}
\centering
\caption{Fit to the complete UV-optical-NIR light curves of SN\,2015bn with the magnetar model in \mosfit\ \citep{guillochon2018}. The fit and derived parameters are essentially unchanged compared to those in \citet{nic2017c}, and naturally account for the new late-time data.} 
\label{fig:fit}
\end{figure}

We model the full light curve of SN\,2015bn using \mosfit\ \citep{guillochon2018}. SN\,2015bn has previously been fit using this code and a magnetar model by \citet{nic2017c}, who describe the methodology. The result is shown in Figure \ref{fig:fit}. Given that new data comprise only $\approx 1\%$ of the total light curve points, it is unsurprising that the fit is unchanged with respect to \citet{nic2017c}. We find a spin period $P/{\rm ms}=2.32\pm0.22$, magnetic field $\log(B/10^{14}{\rm G})=-0.51\pm0.09$, and ejecta mass $\log(M_\mathrm{ej}/M_\odot)=1.04\pm0.03$.
More interesting is that the best fit to the first 400 days gave a reasonably accurate prediction of the evolution at $>1000$ days. The model matches the data at 721 days, and agrees to better than a factor two at 1083 days, though the later data appear systematically above the fit. The previous modelling suggested $\kappa_\gamma\sim 0.01$\cmg, which we confirm here.

\subsubsection{Is there a `missing energy problem'?}

The requirement for inefficient trapping has important implications. Our model implies that by $\sim700$ days the engine is injecting $\sim10^{43}$\,\ergs, but only a few percent is thermalised, indicating a large fraction of `missing energy' escaping. \citet{bhirombhakdi2018} imaged SN\,2015bn in soft X-rays at 725 days. They detected no flux at 0.3--10\,keV to a limit of $\lesssim10^{41}$\,\ergs\, prompting them to conclude that $\lesssim1.5\%$ of the magnetar input escapes in this range.
The energy does not escape in the radio either; using our derived parameters from \mosfit, \citet{margalit2018} predict that the ejecta will remain optically thick to free-free absorption at $\sim 20-40$\,GHz for approximately 10 years, consistent with our VLA non-detections.

\citet{met2014} describe how the magnetar should inflate a nebula of energetic particles and radiation. When the nebula is initially `compact', photon-photon pair creation gives a relatively flat spectral energy distribution (SED) with an upper cut-off at $\sim1-10$\,MeV. Using their equation 13 and our parameters from \mosfit, we find a dimensionless compactness parameter $\ell\lesssim1$ by maximum light and $\ell\sim0.002$ at the timescales we probe here. At low compactness, the SED cut-off moves up to the GeV-TeV range. The dominant opacity is then from photon-matter pair creation, which has an opacity $\kappa_\gamma\sim0.01-0.03$\,\cmg\ over many orders of magnitude in energy \citep{zdziarski1989}. 
The fact that the value of $\kappa_\gamma$ inferred from optical data agrees with this range may provide indirect evidence that the magnetar SED is peaking in high-energy gamma-rays, and that the escape of this radiation is the source of missing energy. 
\citet{renault2018} searched for GeV leakage from SLSNe with \textit{Fermi}, but their limits were not deep enough to detect $\sim10^{43}$\,\ergs, leaving open this possibility.

\subsection{Freeze-out?}

The mechanisms discussed in sections \ref{sec:co}--\ref{sec:mag} assume that energy deposition is instantaneous. However, if the heat source is coupled to the ejecta through ionization and recombination, this assumption holds only if the recombination timescale is shorter than the heating timescale, which may not be true at late times when the ejecta density is low. This process of `freeze-out' can result in a light curve tracking the recombination rate instead of the heating rate \citep{fransson1993,fransson2015}.

Following \citet{kerzendorf2017} and \citet{graur2018a}, we parameterize freezeout as a luminosity source that evolves as $t^{-3}$ (i.e.~in proportion to the density, assuming constant expansion). \citet{graur2018a} define $t_{\rm freeze,50}$ as the time when freezeout accounts for half of the emission. If freezeout dominates by $\sim 700$ days, we find $t_{\rm freeze,50}\sim400$ days. This is much earlier than in SN\,1987A and a number of nearby SNe Ia, for which the timescales are typically $\gtrsim800$ days \citep{fransson1993,graur2018a}. It therefore seems unlikely that freezeout alone can account for the flattening, but more detailed modelling is required here.

\section{Discussion and conclusions}
\label{sec:dis}

We have presented optical imaging, spectroscopy and deep radio limits for SN\,2015bn at $\approx700-1100$ days after maximum light. \hst\ images enabled us to localise the faint SN within its compact host, and reliably extract its flux. We found a significant flattening in the light curve, which is now much slower than \Co\ decay, while the spectrum remains consistent with previous observations at $\sim300$ days.

We showed that the spectrum, colours and decline rate were inconsistent with a light echo. The luminosity, $\sim 10^{41}$\,\ergs, is too large for slowly-decaying radioactive isotopes like $^{57}$Co; the required $\gtrsim 6$\,\M\ far exceeds any physical model of which we are aware. Late-time circumstellar interaction is a more plausible mechanism to slow the light curve, however neither the spectrum nor radio data indicate interaction. In particular, SN\,2015bn lacks the broad \Ha\ seen in other SLSNe that interact at late times \citep{yan2017}.

The light curve shape can be reproduced with a power-law, $\alpha\approx4$, which we show is expected for a magnetar engine with incomplete trapping. In fact, the same magnetar parameters inferred from earlier data naturally predict an evolution in reasonable agreement with our observations.
Our fit suggests that only a few percent of the $\sim10^{43}$\,\ergs\ input is thermalised at this phase, suggesting significant luminosity from leakage at other wavelengths. However, our radio data, and soft X-ray data from \citet{bhirombhakdi2018}, have yielded non-detections. The opacity to magnetar input inferred from our light curve modelling, $\sim0.01$\,\cmg, suggests a harder spectrum, likely concentrated at $\gg10$\,MeV, which may be where the missing energy is escaping. 

While SN\,2015bn is the first SLSN observed to reach a decline much shallower than \Co\ decay, there is a recent example of a SN Ic, iPTF15dtg, exhibiting similar behaviour. \citet{taddia2018} interpreted this as a signature of magnetar powering. We note that the nebular spectrum of iPTF15dtg closely resembles SN\,2015bn, and shows several features, such as prominent \oi\ and [\ion{O}{3}]\,$\lambda$5007, which are more characteristic of SLSNe than normal SNe Ic \citep{mil2013,nic2016c,nic2018}.

The strength of any [\ion{O}{3}] emission in our latest spectrum is difficult to establish given the low S/N, however it is clearly weaker than the line we identify as [\ion{O}{1}]. This is interesting given that \citep{chevalier1992} find that in a pulsar-energised SN at this phase, [\ion{O}{3}] should often be the strongest line. Following their discussion, the high ratio of [\ion{O}{1}]/[\ion{O}{3}] could indicate a large ejecta mass, such that the highly-ionised region does not extend too far in mass coordinate \citep{met2014}, and/or significant clumping (a density enhancement $\gtrsim 10$), which can boost the [\ion{O}{1}] emission \citep[see also][]{jer2017a}.

The latest photometry of SN\,2015bn, at 1068 days, is slightly brighter than the predictions of the basic magnetar model. While we caution that this is based on only two epochs, such an effect could be interpreted as evidence that the power-law is not exactly $\alpha=2$, e.g.~\citet{metzger2018} have shown that accretion onto a magnetar can alter its spin-down. Alternatively, low-level interaction may be a factor, perhaps connected to earlier undulations in the light curve \citep{nic2016b,ins2017}. Finally, we cannot exclude a small contribution from freezeout effects.

Obtaining observations of additional nearby SLSNe at $\gtrsim 500$ days will be required to determine if the slow decline observed in SN\,2015bn is ubiquitous, and whether it is indeed the long-awaited smoking gun for the magnetar. The closest events may hold further promise for detecting leakage of the input energy, and directly probing the engine SED; we suggest such searches should focus on hard X-rays and gamma-rays.

\acknowledgements

We thank Or Graur and Dan Milisavljevic for helpful discussions, Yuri Beletsky for obtaining the LDSS spectrum, and Atish Kamble for his models.
M.N.~is supported by a Royal Astronomical Society Research Fellowship.
Based on observations with the NASA/ESA Hubble Space Telescope. The Space Telescope Science Institute is operated by the Association of Universities for Research in Astronomy, Inc., under NASA contract NAS 5-26555.
This research was funded by grants HST-GO-14743 and HST-GO-15252.
Includes data from the 6.5m Magellan Telescopes at Las Campanas Observatory, Chile.
The National Radio Astronomy Observatory is a facility of the National Science Foundation operated under cooperative agreement by Associated Universities, Inc.
The Berger Time-Domain Group at Harvard is supported in part by the NSF under grant AST-1714498 and by NASA under grant NNX15AE50G. P.K.B.~acknowledges NSFGRP Grant No.~DGE1144152.
R.C.~acknowledges NASA Chandra Grant Award GO7-18046B and NASA XMM-Newton grant 80NSSC18K0665.

\software{MOSFiT, Galfit, Pyraf, SciPy, Matplotlib, SAOImage DS9, HOTPANTS}

\end{document}